# Inertia in the Structure of Space-Time

A. M. Gevorkian[1)] , R. A. Gevorkian*

Institute of Physical Research,

378410, Ashtarak-2, Republic of Armenia

*Institute of Radiophisics and Electronics

378410, Ashtarak-2, Republic of Armenia

**Summary**

In Einstein's equation we suggest a geometrical object substituting the tensor of energy of impulse and tension. The obtained equation, together with the equation for external field, makes up the complete problem of mathematical equations of gravitation, as well as those of inertia. Based on the example of centrally symmetrical field those problems and their consequences are discussed.

Einstein in his work [1] mentions that equation $G_{ik}=0$, where $G_{ik}$ is the Gilbert-Einstein tensor, would be a basis for ideal physical theory, if along with gravitational fields electromagnetic fields are described as well.

However, we will discuss that equation from another standpoint of the general theory. It is part of the unified theory of gravitation and inertia in that part which describes gravitation (external problem), worked out by Einstein (GTR).

The Gilbert-Einstein tensor $G_{ik}$ which is used in gravitational physics is determined as

$$K = \frac{G_{ik}\xi^i\xi^k}{\xi^i\xi_k}, \qquad (1)$$

where K is the mean scalar crookedness of three-dimensional subspace orthogonal to an arbitrary vector $\xi_k$.

---

[1)] e-mail: *dealing51@hotmail.com*

This tensor is connected with the structure of four-dimensional space and plays a fundamental role in the structure of GTR field equation [2].

Let us discuss expression $\frac{\eta_{ik} - g_{ik}}{\xi^i \xi_k}$ determining the value of deformation of four-dimensional Rimman space in relation with Euclidean space.

Let us determine tensor

$$N_{ik} = G_{ik} - \frac{(\eta_{ik} - g_{ik})}{\xi^i \xi_k}, \quad (2)$$

where $\xi_k$ is determined by equation (1), and $\eta_{ik}$, $g_{ik}$ are metrical tensors of Minkowski and Einstein spaces respectively.

Let us introduce tensor $\overline{G}_{ik}$, determined by a different geometry

$$N_{ik} = \overline{G}_{ik} + \frac{(\eta_{ik} - g_{ik})}{\xi^i \xi_k} \quad (3)$$

From equations (2) and (3) we obtain:

$$G_{ik} - \overline{G}_{ik} - 2\frac{(\eta_{ik} - g_{ik})}{\xi^i \xi_k} = 0 \quad (4)$$

When $N_{ik} = 0$ or $G_{ik} = \overline{G}_{ik}$, is derived $\eta_{ik} = g_{ik}$, which represents the case of absolutely flat space-time.

Expression (4) is decomposed by two means into two equivalent equations

$$\overline{G}_{ik} = -2\frac{(\eta_{ik} - g_{ik})}{\xi^i \xi_k}, \quad G_{ik}=0 \quad \text{or} \quad G_{ik} = 2\frac{(\eta_{ik} - g_{ik})}{\xi^i \xi_k}, \quad \overline{G}_{ik} = 0 \quad (5)$$

Thus we can formulate the complete problem in two equivalent ways:

$$\begin{cases} G_{ik} = 2\frac{(\eta_{ik} - g_{ik})}{\xi^i \xi_k} \\ G_{ik} = 0 \end{cases} \quad (6.1, 6.2)$$

$$\begin{cases} \overline{G}_{ik} = -2\dfrac{(\eta_{ik} - g_{ik})}{\xi^i \xi_k} \\ \overline{G}_{ik} = 0 \end{cases} \quad (7.1, 7.2)$$

Equation systems (6) and (7) describe physical phenomena (inertial-gravitational). In these equations the internal problem of gravitation becomes the external problem of inertia, and vice versa.

Equation (6.2) is the external problem of gravitation for the case of centrally symmetrical field with Schwarzschild's solution [3]:

$$ds^2 = \left(1 - \frac{r_0}{r}\right) c^2 dt^2 - r^2 \left(\sin^2\theta \, d\varphi^2 + d\theta^2\right) - \frac{dr^2}{1 - r_0/r} \quad (8)$$

Equation (6.1) describes either the internal problem of gravitation with tensor of energy of impulse and tension, or the geometrical origin, or the external problem of inertia. By writing the tensor equation (6.1) through its components for centrally symmetrical field, we obtain:

$$\begin{cases} \exp(-\lambda)\left(\dfrac{\partial \lambda}{\partial r}\dfrac{1}{r} + \dfrac{1}{r^2}\right) - \dfrac{1}{r^2} = 0 \\ \dfrac{\partial \lambda}{\partial t} = 0 \end{cases} \quad \text{where} \quad \begin{cases} g_{00} = \exp(-\lambda) \\ g_{33} = \exp(\lambda) \end{cases} \quad (9)$$

with solution $g_{00} = \dfrac{1-r}{r_0}$ and the respective metrics.

$$ds^2 = \left(1 - \frac{r}{r_0}\right) c^2 dt^2 - r^2 \left(\sin^2\theta \, d\varphi^2 + d\theta^2\right) - \frac{dr^2}{1 - r/r_0} \quad (10)$$

Here $r_0$ is the constant of the discussed metrics, which will be determined the same way as it is in the Schwarzschild solution.

In expression (10) we will discuss two cases:

1. For weak field we have: (Newton-Mach approximation)

$$ds^2 = \left(1 - \frac{r}{r_0}\right) c^2 dt^2$$

if $r_0 = r_k = \dfrac{h}{Mc^2}$ , then for force we obtain:

$$F = \dfrac{Mmc^3}{h} \dfrac{\vec{r}_0}{|r_0|} \qquad (11)$$

by analogy with Newton's law of terrestrial gravitation obtained in GTR. In expression (11) the force does not depend on distance, which means that all members (stars) of the Metagalaxy have equal parts in the formation of this force.

Due to symmetrical distribution of stars in an arbitrary point of space, the resulting affect of this force is equal to zero F=0. This is Newton's first law. The law of inertia in the real physical space is obtained, which substitutes absolute space-time [4]. Any non-uniform motion is connected with non-inertial system of reference in relation to the Metagalaxy. Such motion distorts the local isotropness, and therefore the summary affect of the Space on this body is equal to zero. Forces of inertia are formed, such as centrifugal and Coriolis forces, and others. It is considered that such forces are formed exclusively due to the choice of non-inertial system of reference. In Newton's mechanics these forces are introduced formally, allowing to apply Newtonian laws of motion, even though in fact they are not fulfilled. For example, if rotations of bodies are real in respect to static stars, then centrifugal forces are also real, which have a global cosmic origin. All this is in tune with Mach's theory, according to which inertial systems are systems in respect to the stars, which have no acceleration, in other words in respect to some mean distribution of matter in the Space [5]. Moreover, a body has inertial properties only because there are other bodies in the Space.

Existence of isotropic Space is sufficient for constructing inertial systems. Expression (11) together with the law of terrestrial gravitation obtained from Schwarzschild's solution form the approximation of weak field (Newton-Mach).

2. In particular case if in formula (10)

$$ds^2 = r^2\left(\sin^2\theta d\varphi^2 + d\theta^2\right) \qquad (12)$$

then it is equivalent to

$$\left(1 - \dfrac{r}{r_0}\right)c^2 dt - \dfrac{1}{1 - r/r_0} dr = 0 \qquad \text{or}$$

$$r = r_0(1 - \cos\theta), \qquad r = \frac{h}{mc}(1 - \cos\theta) \qquad (13)$$

Compton's formula is received for diffusion of light on charged particles, where $\frac{v}{c} = \cos\theta$, $r_0 = \frac{h}{mc}$ is the Compton or inertial radius.

In Schwarzschild's solution this particular case brings us to Einstein's formula [6], which was obtained even before creation of GTR.

$$c = c_0(1 - 2GM/c^2 r) \qquad (14)$$

Let us call particular cases when formulas (13) and (14) are received, cases of Einstein-Compton. Here m is active inert mass. The observed Compton effect on charged particles is the exhibition of active inert mass in metrics (10). For large m this effect is not observed because of its small value.

Schwarzschild's solution and metrics (10) do not coincide anywhere else except in the point

$$\frac{r_{gr}}{r} = \frac{r}{r_k} \qquad r^2 = r_{gr} r_k \qquad \text{or} \qquad r^2 = \frac{Gm_{gr}}{c^2} \frac{h}{cm_{in}}$$

If the active gravitational mass is equal to the active inert mass (principle of equivalence in the wider sense), then we will have for radius

$$r_0^2 = \frac{Gh}{c^3} \sim 10^{-66} \qquad (15)$$

As a result both metrics are written through the same formula:

$$ds^2 = \left(1 - \frac{m}{m_0}\right) c^2 dt^2 + r_0^2 (\sin^2\theta \, d\varphi + d\theta^2) - \frac{dr^2}{1 - m/m_0} \qquad (16)$$

and represents metrics describing real physical space, where

$$m_0 = \left(\frac{hc}{G}\right)^{1/2}, \qquad r_0 = \left(\frac{Gh}{c^3}\right)^{1/3}$$

Analysis of formula (16) is still to be done. However the following can be concluded here:

1. In this metrics when m=0, we have Minkowski's metrics - empty, absolute mathematical space. Therefore in order for the tensor of crookedness to be equal to zero, existence of empty space m=0 is necessary and sufficient.
2. Physical real space is by no means Minkowski's empty space, and is viewed as physical space of Rimman-Einstein-Mach.
3. $m=m_0$ or $r=r_0$ is a special point - radius of inversion. In this point ($r=r_{gr}=r_{in}=r_0$) both metrics either have a same kind of gap, or the physical metrics in the point is not determined.

It is known that if the matter is under gravitational radius, then we do not receive any other information about that matter except for the gravitational field. In this case the matter is under gravitational radius, as well as under inertia radius. Such bodies will no longer interact with each other. Such particles can be used to construct physical vacuum.

**Calculation of the Mass**

From equation

$$G_{ik} = 2\frac{(\eta_{ik} - g_{ik})}{\xi^i \xi_k}, \text{ where tensor of energy of impulse and tension}$$

$$2\frac{(\eta_{ik} - g_{ik})}{\xi_i \xi^k} = T_{ik}.$$

For centrally symmetric field we can write:

$$M = \int T_{00} r^2 dr \qquad (17)$$

where $T_{00} = 2\frac{(\eta_{00} - g_{00})}{r^2}$

Substituting the value of $g_{00}$ in formula (17) and integrating it, we obtain

$$M = \int_x^{1/x} \frac{r^2}{rr_0} dr = \frac{1}{2}\left(x^2 - \frac{1}{x^2}\right) \qquad (18)$$

The internal solution of $G_{ik}$ is at the same time the external solution for $\overline{G_{ik}}$; therefore we may calculate

$$\overline{M} = \int_{1/x}^{x} \frac{2r_0 r^2}{r^3} dr = 2\left(\ln x - \ln \frac{1}{x}\right) = 2\ln|x^2| \qquad (19)$$

Due to equivalence of mass $M = \overline{M}$, in the wide sense we can obtain a transcendental algebraic equation:

$$x^4 - 1 = 4|x^2|\ln|x^2| \qquad (20)$$

roots of which are $|x_0^2| = 1$, $x_1 = x_0 \alpha^{1/4}$, and their reverse values.

It is assumed that x is connected with the constant of fine structure α, or is the value of magnetic charge:

$$x_5^2 = \frac{\alpha}{\sqrt{2}} \qquad (21)$$

If expression (21) determines the value of electric charge, then x can be considered to be the value of so called hyperelectric charge, which can be used in further works of electrodynamics.

The following is the picture we receive: all types of energy are sources of the gravitational field, except the energy of the gravitational field itself. The source of inert field is the mass, caused only by equivalent mass of energy of gravitational field. And if the principle of equivalence in the wide sense is in place, in other words the active gravitational mass is equal to active inert mass, then we receive equation (20), which in our opinion is a fundamental one.

## Conclusion

1. It is suggested to substitute the physical tensor of energy of impulse and tension with mathematical (geometrical) tensor in the right side of Einstein equation.
2. Along with the external solution of Schwarzschild a new metrics is received, which corresponds to the internal solution (for centrally symmetrical field).
3. In the approximation of the weak field (Newton, Mach) mathematical expressions are received, which are in tune with Mach's principle and are connected with the phenomenon of inertia.
4. Compton's formula is received (the case of Einstein-Compton in the article).

5. Using the principle of equivalence for sources of filed (of gravitation and inertia), the value of magnetic charge, or the constant of fine structure is received.